\documentclass[12pt]{iopart}
\usepackage{iopams}
\usepackage{graphicx,multirow}

\begin{document}

\title{Role of electronic excitations in magneto-Raman spectra of graphene}
\author{Oleksiy Kashuba}
\address{Institute for Theory of Statistical Physics, RWTH Aachen, 52056 Aachen, Germany}
\ead{kashuba@physik.rwth-aachen.de}
\author{Vladimir I. Fal'ko}
\address{Department of Physics, Lancaster University, Lancaster, LA1~4YB, UK}

\begin{abstract}
We investigate the signature of the low-energy electronic excitations in the Raman spectrum of monolayer and bilayer graphenes.
The dominant contribution to the Raman spectra is due to the interband electron-hole pairs, which belong to the irreducible representation A$_{2}$ of the point group C$_{6v}$ of the graphene lattice, and are characterised by crossed polarisation of incoming and outgoing photons.
At high magnetic fields, this is manifested by the excitation of electron-hole (e-h) inter-Landau-level transitions with selection rule $n^{-}\to n^{+}$.
Weaker Raman-active inter-Landau-level modes also exist.
One of those has a selection rule similar to the infrared absorption process, $n^{-}\to(n\pm1)^{+}$, but the created e-h excitation belongs to the irreducible representation $E_{2}$ (rather than $E_{1}$) and couples to the optical phonon mode, thus undergoing an anticrossing with the optical phonon G-line in Raman in strong magnetic field.
The fine structure of acquired by the G-line due to such anticrossing depends on the carrier density, inhomogeneity of doping, and presence of inhomogeneous strain in the sample.
\end{abstract}

\maketitle

\section{Introduction}

Experimental observation of electronic excitations in graphene is mostly associated with angle-resolved photoemission spectroscopy \cite{Ohta, Mucha1} and optical absorption in monolayers~\cite{Potemski2,Kim2,Kuzmenko1,Basov,AbergelFalko,Geim,AbergelRussellFalko} and bilayers~\cite{Basov, Kuzmenko1, Kuzmenko2, Kim1, Kim3, Heinz, Kuzmenko3,GeimNovoselov3, Gaskell} where Raman spectroscopy has been used to study graphene phonons~\cite{Ferrari}, and became the method of choice for determining the number of atomic layers in graphitic flakes~\cite{Ferrari,Graf,CastroNeto1,Jiang,Potemski1,Berciaud,BalandinCalizo}.
In particular, single- and multiple-phonon-emission lines in the Raman spectrum of graphene and the influence of the electron-phonon coupling on the phonon spectrum have been investigated in great detail~\cite{GeimNovoselov1,CastroNeto2,Basko1_1,Basko1_2,Ando,Kechedzhi1,Pinczuk1,Pinczuk2,FerrariBasko}.

This manuscript presents a theory of inelastic light scattering in the visible range of photon energies accompanied by electronic excitations in mono- and bilayer graphene both with and without an external magnetic field.
We classify the relevant modes according to their symmetry, analyse peculiar selection rules for the Raman-active excitations of electrons between Landau levels in graphene at strong magnetic fields, review the fine structure of the phonon-related Raman under conditions of magneto-phonon resonance (when the inter-Landau-level transition is in resonance with phonon), and model the dependence of the spectra on the value and homogeneity of carrier density and strain in the sample.

Graphene is a gapless semiconductor~\cite{Wallace,RMPgraphene,GeimNovoselov2,McCann2}, with an almost linear Dirac-type spectrum $\varepsilon =\alpha vp$ in monolayer case and parabolic $\varepsilon =\alpha p^{2}/2m$ for bilayer, in the conduction ($\alpha =+$) and valence ($\alpha =-$) band, which touch each other in the corners of the hexagonal Brillouin zone, usually called valleys.
In bilayer graphene there are two additional parabolic band with the gap $2\gamma_{1}$.
The conduction-valence band degeneracy in the Brillouin zone corners of the band structure of graphene is prescribed by the hexagonal symmetry C$_{6v}$ of its honeycomb lattice, and it is natural to relate Raman-active modes to the irreducible representations of the point group C$_{6v}$.

We argue that the dominant electronic modes generated by inelastic scattering of photons with energy $\Omega$ less than the bandwidth of graphene are superpositions of the interband electron-hole pairs which have symmetry of the representation A$_{2}$ of the group C$_{6v}$ and are odd with respect to the inversion of time.
Their excitation process may go in two ways both of which consist of the same steps, but in reversed order.
In the first process the absorption of a photon with energy $\Omega$ transfers an electron from an occupied state in the valence band into a virtual state in the conduction band, followed by emission of the second photon with energy $\tilde{\Omega}=\Omega-\omega$.
In the second process the reversed order of events leads the the virtual electron state with large deficit of energy as opposed to the first process with large excess.
Net amplitude is determined by the sum of partial amplitudes of the opposite processes.
The dominance of such process over the one involving the contact interaction~\cite{PlatzmanWolff,PlatzmanWolff2,Chinese1} of an electron with two photons is a peculiarity of the Dirac-type electrons in graphene.

Strong magnetic field leads to the quantisation of the electronic spectrum into Landau levels~\cite{McClure}.
In graphene Landau levels (LL) appear symmetrically in the conduction bands $\epsilon^{\alpha}_{n}$ ($\alpha=\pm$) with peculiar zero-energy LLs in monolayer ($n=0$) and bilayer ($n=0.1$).
For the inelastic light scattering in graphene, we find peculiar selection rules, $n^{-}\rightarrow n^{+}$ of the dominant Raman-active transitions (solid line in Fig.~\ref{fig:graman}), in contrast to the $\Delta n=\pm 1$ transitions between Landau levels which are dominant in the absorption of left and right-handed circularly polarised infrared photons~\cite{AbergelFalko}.
The weaker Raman mode with a similar inter-LL selection rule is also differs from absorption process, having different symmetry properties which allow the interaction with Raman scattering of light on optical phonons creating anticrossings at the G-peak line, see Fig.~\ref{fig:magscheme}.
Raman spectroscopy, therefore, provides data supplementary to that obtained in optical absorption.

The following theory is based upon the effective low-energy Hamiltonian derived from the tight-binding model of electron states in graphene.
It describes electrons in the conduction and valence bands around the Brillouin zone (BZ) corners $K$ ($\xi=+$) and $K'$ ($\xi=-$)~\cite{Dresselhaus,McCann1}.
For monolayer graphene the effective Hamiltonian reads,
\begin{equation}
H_{m}=\xi v\boldsymbol{\sigma }\cdot \mathbf{p}
-
\frac{v^{2}}{6\gamma_{0}}\bigl(\sigma ^{x}(p_{x}^{2}-p_{y}^{2})-2\sigma^{y}p_{x}p_{y}\bigr).
\label{Hm}
\end{equation}
Here $\boldsymbol{\sigma}=(\sigma^{x},\sigma^{y})$ are Pauli matrices acting on the sublattice components of electronic states on A and B sublattices of graphene honeycomb lattice, and $\mathbf{p}$ being the in-plane momentum counted from the BZ corner.
The first term in Eq.~\eref{Hm} determines Dirac spectrum electrons.
The monolayer hopping parameter $\gamma_{0}\approx 3$eV determines the bandwidth, $\sim 6\gamma _{0}$, and Dirac velocity of electrons $v\approx 10^{8}$cm/s~\cite{Kuzmenko2, Kim3, CastroNeto1, Malard2}.
The second term takes into account weak trigonal warping~\cite{Dresselhaus} breaking  continuous rotational symmetry of Dirac Hamiltonian.
The basis is constructed of wavefunctions corresponding to atomic sites $A,B$ in the valley $K$, and $B,A$ in $K'$.
Since 4-spinors $\{\varphi_{AK},\varphi_{BK},\varphi_{BK'},\varphi_{AK'}\}$ realise 4D irreducible representation of the full symmetry group of the crystal, matrix operators can be combined into irreducible representations \cite{Kechedzhi2,Basko1_1,Basko1_2} of the group C$_{6v}$, see Table~\ref{tab:Reps}.

\begin{table}[tbp]
\centering%
\begin{tabular}{|c|c|c|c|c|c|c|}
\hline
C$_{6v}$ rep. & A$_{1}$ & A$_{2}$ & B$_{1}$ & B$_{2}$ & E$_{1}$ & E$_{2}$ \\ 
\hline
matrix & $1$ & $\sigma^{z}$ & $\xi$ & $\xi\sigma^{z}$ & $\xi\boldsymbol{\sigma }$ & $\boldsymbol{e}_{z} \times \boldsymbol{\sigma}$
\\
\hline
$t\to -t$ & $+$ & $-$ & $-$ & $+$ & $-$ & $+$
\\
\hline
 & $(\mathbf{l}\cdot \mathbf{\tilde{l}}^{*})$ & $(\mathbf{l}\times \mathbf{\tilde{l}}^{*})_{z}$ & & & $\mathbf{l}, \mathbf{\tilde{l}}^{*}$ & $\mathbf{M},\mathbf{d}$
\\
 & $p^{2}$ & & & & $\mathbf{p}$ & $(2p_{x}p_{y},p_{x}^{2}-p_{y}^{2})$
\\
 & & & & & & $\mathbf{u}=\frac{1}{\sqrt 2}(\mathbf{u}_{A}-\mathbf{u}_{B})$
\\
\hline
\end{tabular}%
\caption{Classification of the Hamiltonian components by C$_{6v}$ irreducible representations.}
\label{tab:Reps}
\end{table}

\begin{figure}[tbp]
\centering
\includegraphics[width=.9\columnwidth]{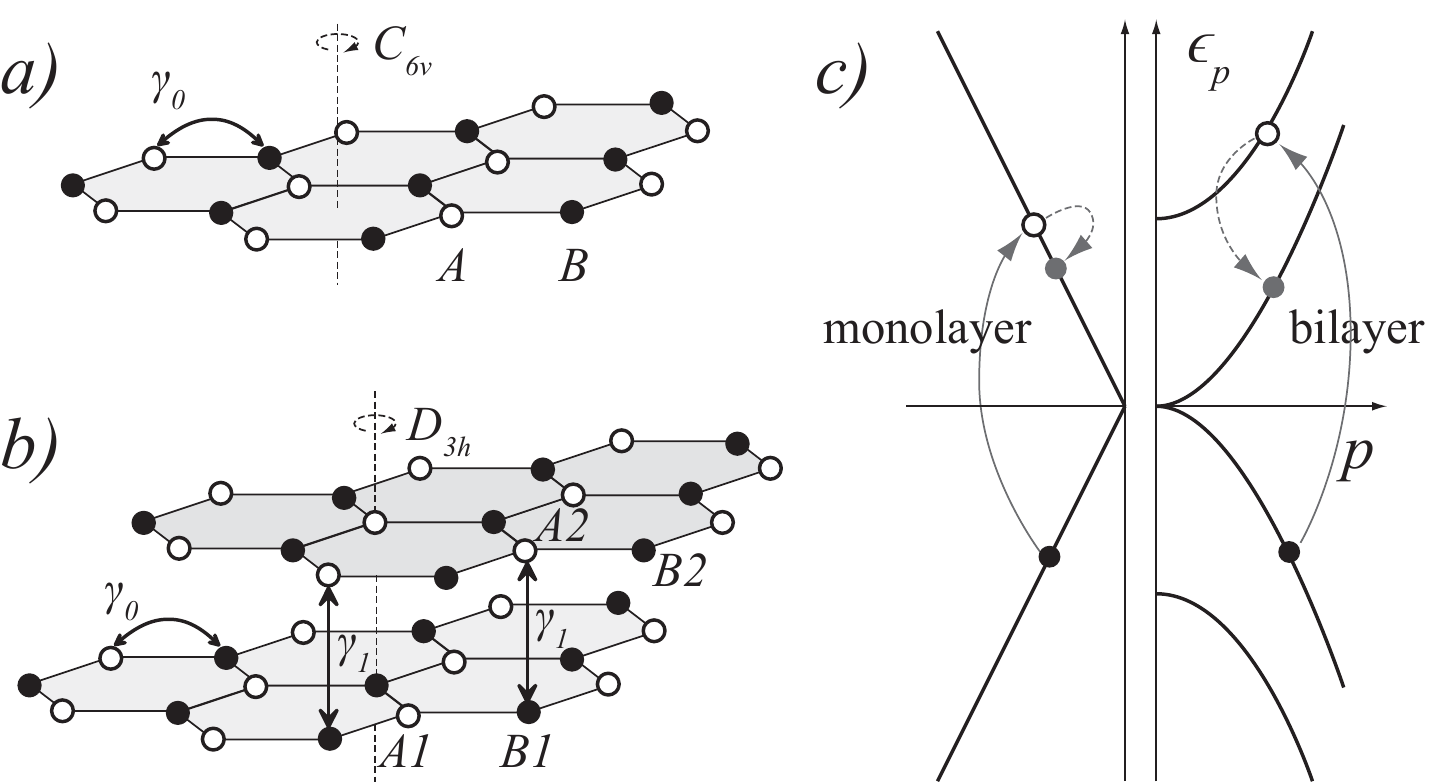}
\caption{%
Schematic of the monolayer (a) and bilayer (b) graphene crystal lattice.
(c) The band structure of mono- and bilayer graphene in the vicinity of the $K$ point along the $p_{x}$ axis.
Also shown are some of the two-step processes leading to the creation of an electron-hole pair in the low-energy bands accompanied by the absorption of a photon followed by emission.
Note that the energy of any intermediate state (white circle) is much smaller than photons energy $\Omega$, $\tilde{\Omega}$.}
\label{fig:lattice}
\end{figure}

Bilayer graphene, Fig.~\ref{fig:lattice}b, consists of two coupled sheets of monolayer with $AB$ (Bernal) stacking (as in bulk graphite~\cite{Dresselhaus2}), and has symmetry $D_{3h}$ which differs from $C_{6v}$ by accompaniment of a $z\to-z$ reflection to the rotations by $\pi/3$, comparing to monolayer graphene, but, otherwise, $D_{3h}$ and $C_{6v}$ are isomorphic.
Its unit cell contains four inequivalent atoms $A1$, $B1$, $A2$ and $B2$ where numbers denote the layer.
The conventional Hamiltonian in the vicinity of the valleys is 
\begin{equation}
H_{b}=
\left(
\begin{array}{cc}
\frac{v_{3}}{v} \sigma^{x} H_{m} \sigma^{x} & H_{m} \\
H_{m} & \gamma_{1}\sigma^{x}
\end{array}
\right),
\label{4b_Hamiltonian}
\end{equation}
where, $v_{3}\sim 0.1 v$ is related to the weak direct $A1\to B2$ interlayer hops~\cite{McCann2, CastroNeto1, Malard2}, and $\gamma_{1}\sim 0.4$eV is the direct interlayer coupling~\cite{Ohta, Basov, Kuzmenko1, Kuzmenko2, Kim3, CastroNeto1, Pinczuk2, Chakraborty, Malard2}.
The basis is constructed using components corresponding to atomic sites $A1,B2,A2,B1$ in the valley $K$ and $B2,A1,B1,A2$ in $K'$.

The electronic dispersions around the $K$ point for mono- and bilayer are compared in Fig.~\ref{fig:lattice}.
Two low-energy bands touch each other at the centre of the valley, which is also the position of the Fermi energy in the neutral structure (a neutrality point), and are described by the effective two-band low-energy Hamiltonian written in the basis of orbitals on the sites $A1$ and $B2$~\cite{McCann2},
\begin{equation}
\label{2b_Hamiltonian}
{\hat{H}}_{b\,\textrm{eff}}  = -\frac{v^{2}}{\gamma_{1}}\bigl(\sigma^{x}( p_{x}^{2} - p_{y}^{2}) + 2 \sigma^{y} p_{x}p_{y} \bigr).
\end{equation}
Two other, referred to as high-energy bands, are split by the interlayer coupling $\gamma_{1}$ from the neutrality point.

\section{Inelastic light scattering in graphene at zero magnetic field}

To include the interaction of the electrons with photons in the effective Hamiltonians~\eref{Hm} and~\eref{4b_Hamiltonian}, we construct the canonical momentum $\mathbf{p}-\frac{e}{c}\mathbf{A}$, where the vector potential of light
\begin{equation}
\mathbf{A}=
\sum_{\mathbf{l},\mathbf{q},q_{z}}
\frac{\hbar c}{\sqrt{2\Omega}}
\left(\mathbf{l}\,e^{i(\mathbf{q}\mathbf{r}-\Omega t)/\hbar }b_{\mathbf{q},q_{z},\mathbf{l}}+h.c.\right),
\label{eq:Adef}
\end{equation}
includes an annihilation operator $b_{\mathbf{q},q_{z},\mathbf{l}}$ for a photon with in-plane momentum $\mathbf{q}$, energy $\Omega$ (which determines its out-of-plane momentum component $q_{z}=\sqrt{\Omega^{2}/c^{2}-\mathbf{q}^{2}}$), and polarisation $\mathbf{l}$.
Expanding the Hamiltonians up to the second order in the vector potential one obtains the interaction part,
\begin{equation}
\label{int_Hamiltonian}
H_{\textrm{int}} = -\frac{ev}{c}\mathbf{J}\cdot\mathbf{A} + \frac{e^{2}}{2c^{2}}\sum_{i,j}(\partial^{2}_{p_{i} p_{j}}H) \, A_{i}A_{j},
\end{equation}
where $(ev/c) J_{i}=(e/c)\partial_{p_{i}}\hat{H}$ is the current vertex and $\frac{e^{2}}{2c^{2}} \partial^{2}_{p_{i} p_{j}}H$ is the two-photon contact-interaction tensor.

\begin{figure}[tbp]
\centering
\includegraphics[width=\columnwidth]{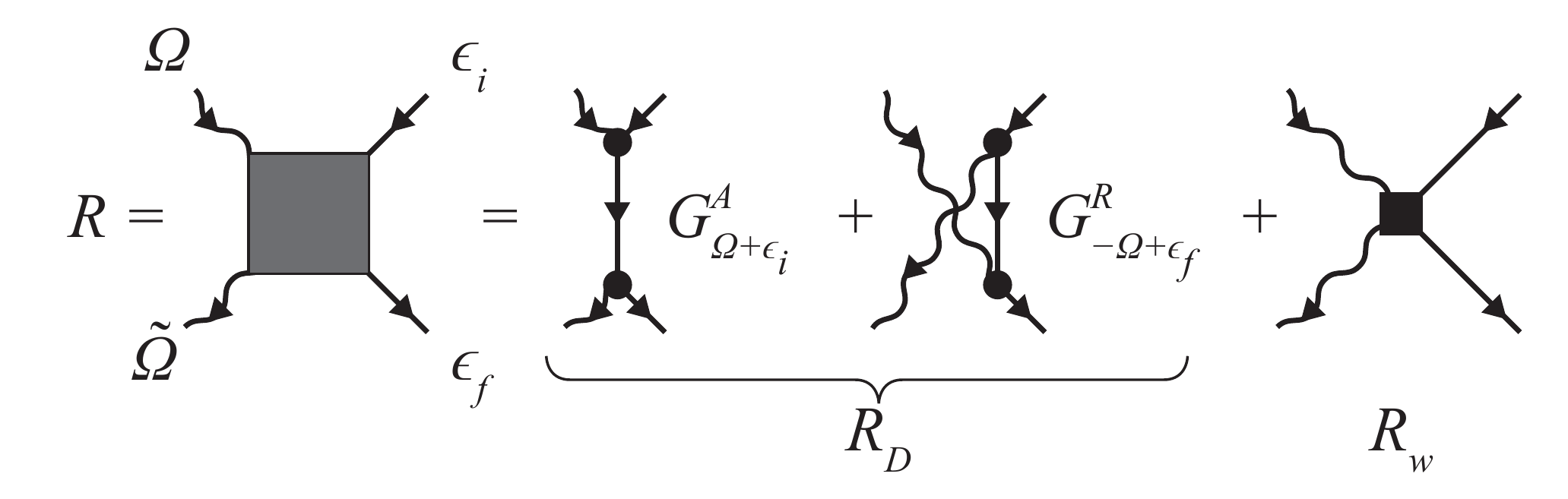}
\caption[Raman diagrams]{Feynman diagrams for Raman scattering amplitude $R$ from ground state to final state with electron-hole excitation.}
\label{fig:diagrams}
\end{figure}

The full amplitude $R=R_{D}+R_{w}$ of inelastic Raman scattering of the photon on electrons changing its energy $\Omega$ to $\tilde{\Omega} = \Omega - \omega$ ($\omega=\epsilon_{f}-\epsilon_{i}$ is the Raman shift), and momentum and polarisation from $\mathbf{q},\mathbf{l}$ to $\tilde{\mathbf{q}},\mathbf{\tilde{l}}$ is described by the Feynman diagrams shown in Fig.~\ref{fig:diagrams}.
The building blocks of the diagrams include Green's functions for the electrons
and the electron-photon interaction vertices: 
\begin{eqnarray}
\raisebox{-12pt}{\begin{picture}(28,28)(-14,-14)
\thicklines
\put(-10,0){\line(1,0){20}}
\put(0,0){\vector(1,0){4}}
\end{picture}} &=&G_{\varepsilon}^{R(A)}=(\varepsilon-H\pm i0)^{-1},  \nonumber \\
\raisebox{-12pt}{\begin{picture}(28,28)(-14,-14)
\thicklines
\put(-8,12){\oval(8,8)[bl]}
\put(-8,4){\oval(8,8)[tr]}
\put(0,4){\oval(8,8)[bl]}
\put(-8,8){\vector(1,-1){4}}
\put(0,0){\line(1,1){12}}
\put(8,8){\vector(-1,-1){4}}
\put(0,0){\line(1,-1){12}}
\put(5,-5){\vector(1,-1){4}}
\put(0,0){\circle*{4}}
\end{picture}} &=&\frac{ev\hbar }{\sqrt{2\Omega}}\mathbf{J}_{\mathbf{q}}\cdot \mathbf{l},\qquad 
\raisebox{-12pt}{\begin{picture}(28,28)(-14,-14)
\thicklines
\put(-8,-12){\oval(8,8)[tl]}
\put(-8,-4){\oval(8,8)[br]}
\put(0,-4){\oval(8,8)[tl]}
\put(-4,-5){\vector(-1,-1){4}}
\put(0,0){\line(1,1){12}}
\put(8,8){\vector(-1,-1){4}}
\put(0,0){\line(1,-1){12}}
\put(5,-5){\vector(1,-1){4}}
\put(0,0){\circle*{4}}
\end{picture}}=\frac{ev\hbar }{\sqrt{2\tilde\Omega}}\mathbf{J}_{-\tilde{\mathbf{q}}}\cdot \mathbf{\tilde{l}}^{*}, \label{eq:diagramelem} \\
\raisebox{-12pt}{\begin{picture}(28,28)(-14,-14)
\thicklines
\put(-8,-12){\oval(8,8)[tl]}
\put(-8,-4){\oval(8,8)[br]}
\put(0,-4){\oval(8,8)[tl]}
\put(-4,-5){\vector(-1,-1){4}}
\put(-8,12){\oval(8,8)[bl]}
\put(-8,4){\oval(8,8)[tr]}
\put(0,4){\oval(8,8)[bl]}
\put(-8,8){\vector(1,-1){4}}
\put(0,0){\line(1,1){12}}
\put(8,8){\vector(-1,-1){4}}
\put(0,0){\line(1,-1){12}}
\put(5,-5){\vector(1,-1){4}}
\multiput(-3,-3)(.5,0){12}{\line(0,1){6}}
\end{picture}}&=&
R_{w}=\frac{e^{2}\hbar^{2}}{2\sqrt{\Omega\tilde\Omega}} (\partial^{2}_{p_{i}p_{j}}H) l_{i} \tilde{l}_{j}^{*} \delta_{\mathbf{p,p+q-\tilde{q}}}. \nonumber
\end{eqnarray}
Thus, the inelastic light scattering amplitude is contributed by a one-step process $R_{w}$ (contact interaction~\cite{PlatzmanWolff2}) and a two-step process $R_{D}$ involving an intermediate virtual state.
The two-step process, such as shown in Fig.~\ref{fig:diagrams}, consists of an absorption (emission) of a photon with energy $\Omega$ ($\tilde{\Omega}$) transferring an electron with momentum $\mathbf{p}$ from an occupied state in the valence band into a virtual intermediate state, followed by another electron emission (absorption) of the second photon with energy $\tilde{\Omega}$ ($\Omega$) which moves the electron to the final state with momentum $\mathbf{p+q-\tilde{q}}$,
\begin{equation}
R_{D}=\frac{e^{2}\hbar^{2}v^{2}}{2\sqrt{\Omega\tilde\Omega}}\Bigl(
(\mathbf{J}_{\mathbf{q}}\cdot \mathbf{l}) G^{A}_{\Omega+\epsilon_{i}} (\mathbf{J}_{-\tilde{\mathbf{q}}}\cdot \mathbf{\tilde{l}}^{*})
+
(\mathbf{J}_{-\tilde{\mathbf{q}}}\cdot \mathbf{\tilde{l}}^{*}) G^{A}_{-\Omega+\epsilon_{f}} (\mathbf{J}_{\mathbf{q}}\cdot \mathbf{l})
\Bigr).
\label{eq:RDdef}
\end{equation}

In graphene, the two-step process via the virtual state and the contact interaction process have different properties and polarisation selection rules, which reflects their intricate relation with the irreducible representations of symmetry group of the crystal.
Generally, the components of the scattering amplitude $R^{ij}$ realise a representation of the symmetry group $C_{6v}$.
Since vectors $\mathbf{l}$, $\tilde{\mathbf{l}}^{*}$ belong to $E_{1}$ representation, the $R$ representation can be expanded in irreducible ones
\begin{equation}
E_{1}\otimes E_{1}=A_{1}\oplus A_{2}\oplus E_{2}.
\end{equation}
Forming the corresponding combinations from the components of the polarisation vector, see Table~\ref{tab:Reps}
\begin{eqnarray}
\Xi_{s}=\bigl|\mathbf{l}\times \mathbf{\tilde{l}}^{*}\bigr|^{2},
\qquad
\Xi_{s2}=\bigl|\mathbf{l}\cdot \mathbf{\tilde{l}}^{*}\bigr|^{2},
\qquad
\Xi_{o}=|\mathbf{d}|^{2}=1+(\mathbf{l}\times \mathbf{l}^{*})(\mathbf{\tilde{l}}\times \mathbf{\tilde{l}}^{*}),
\\
\mathrm{where}\quad \mathbf{d}=(l_{x}\tilde{l}_{y}^{*}+l_{y}\tilde{l}_{x}^{*},l_{x}\tilde{l}_{x}^{*}-l_{y}\tilde{l}_{y}^{*}),
\quad
\mathrm{and}\quad \Xi_{s}+\Xi_{s2}+\Xi_{o}=2,
\end{eqnarray}
we can write the scattering probability in the most general form
\begin{equation}
w=w_{s}\Xi_{s}+w_{s2}\Xi_{s2}+w_{o}\Xi_{o}.
\end{equation}
The first two terms with polarisation factor $\Xi_{s}$ and $\Xi_{s2}$, corresponding to $A_{2}$ and $A_{1}$ representations respectively, describe the contribution of photons scattered with the same circular polarisation as the incoming beam.
The third term, with polarisation factor $\Xi_{o}$, corresponding to $E_{2}$ representation, represents the scattered photons with circular polarisation opposite to the incoming beam.

Microscopically, the probability for a photon to undergo inelastic scattering from the state $(\mathbf{q},q_{z})$ with energy $\Omega $ into a state $\mathbf{(\tilde{q}},\tilde{q}_{z})$ with energy $\tilde{\Omega}$, by exciting an e-h pair in graphene, is 
\begin{equation}
w=\frac{2}{\pi\hbar^{3}}\sum_{\alpha_{i}\,\alpha_{f}}\int d^{2}\mathbf{p} |\langle f|R|i\rangle|^{2} \times f_{i}\left( 1-f_{f} \right)\delta\left( \epsilon_{i} + \omega -\epsilon_{f} \right), \nonumber
\end{equation}
where $f_{i}$ and $f_{f}$ are filling factors of the initial and final electronic state, respectively, the spin and valley degeneracies have already been taken into account, and initial and final states are defined by the momentum $\mathbf{p}$ and (sub)band index $\alpha_{i/f}$: $|i(f)\rangle=|\mathbf{p}(\mathbf{p+q-\tilde{q}},\alpha_{i/f})\rangle$.
The angle-integrated spectral density of Raman scattering $g(\omega)$ is
\begin{equation}
g(\omega)  = \frac{1}{c}\int\int \frac{d^{2}\tilde{\mathbf{q}}d\tilde{q}_{z}}{(2\pi\hbar)^3}\, w\,\, 
\delta \left( \tilde{\Omega} - c\sqrt{\tilde{\mathbf{q}}^{2}+{\tilde{q}_{z}}^{2}} \right),
\label{spectral_density}
\end{equation}
and quantum efficiency $I=\int g d\omega$ expresses the total probability for single incoming photon to scatter inelastically on an electron and excite an electron-hole pair in the low-energy part of the spectrum.

\subsection{Raman scattering in monolayer graphene}

\label{sec:ramanmonolayer}

The calculation of the Raman spectrum and intensity of the signal in each of the distinct polarisation components, $w_{s}$, $w_{s2}$, and $w_{o}$ requires the explicit expression for the Raman scattering amplitude $R$.
Raman measurements are usually performed at $\Omega,\tilde{\Omega}\sim 1\div3$\,eV, which is much smaller than the bandwidth~\cite{Ferrari}.
Raman shift, which measures the electronic excitation energy, is $\omega\lesssim0.2$\,eV allowing us to expand the Green's functions over the large $\Omega$, and to perform summation over the intermediate virtual states of the process.

Both the electron and the hole of the e-h excitation have almost the same momentum ($\mathbf{p}+\mathbf{q}-\tilde{\mathbf{q}}$ and $\mathbf{p}$, respectively), since $\mathbf{q},\tilde{\mathbf{q}}\ll\mathbf{p}$ and the momentum transfer from light is negligible ($v/c\sim 3\cdot 10^{-3}$), so that below we use $\mathbf{J}_{\mathbf{q}}=\xi\boldsymbol{\sigma} \delta_{\mathbf{p,p+q}}$.
For $vp\ll\Omega$ the dominant contributions to the Raman scattering amplitude are
\begin{eqnarray}
R_{D} &\approx
\frac{(e\hbar v)^{2}}{\Omega^{2}}\left(-i \sigma^{z} (\mathbf{l}\times \mathbf{\tilde{l}}^{*})_{z}+ \frac{\mathbf{M}\cdot\mathbf{d}}{\Omega}\right),
\\
\mathbf{M}&=\xi v (\sigma_{x}p_{y}+\sigma_{y}p_{x},\sigma_{x}p_{x}-\sigma_{y}p_{y}),\nonumber
\\
R_{w} &=\frac{e^{2}v^{2}\hbar ^{2}}{6\Omega \gamma_{0}}(\mathbf{e}_{z}\times \boldsymbol{\sigma })\cdot \mathbf{d},
\label{eq:RD}
\end{eqnarray}
where the main order of $R_{D}$ has the matrix form of representation $A_{2}$ of $C_{6v}$.
Contact interaction $R_{w}$ is responsible for the creation of the excitations with the symmetry of the representation $E_{2}$, see Table~\ref{tab:Reps}.
For scattering of photons with $\Omega <\gamma _{0}$, $R_{w}\ll $ $R_{D}$~\cite{Kashuba}.

In undoped graphene the inter-band e-h pairs with $\epsilon_{f}\approx-\epsilon_{i}\approx\omega/2$ are the only allowed electronic excitations.
The probability of the creating an electronic excitation by photons with $\Omega <\gamma_{0}$, $\mathbf{q}-\tilde{\mathbf{q}}\ll\mathbf{p}$ ($\omega\gg \frac{v}{c}\Omega$)
in the main order of the expansion in $\omega/\Omega\ll 1$ is
\begin{equation}
w\approx
\hbar e^{4}v^{2}\frac{\omega}{\Omega^{4}}\left\{
\Xi_{s}+\frac{\Omega^{2}}{2(6\gamma_{0})^{2}}\Xi_{o}
\right\}.
\label{W(0)}
\end{equation}
It is dominated by the contribution $R_{D}$ of the first two diagrams in Fig.~\ref{fig:diagrams}, describing two-step processes involving a virtual intermediate state.

Doping in graphene, with chemical potential $\mu \gg \Omega v/c$ blocks inter-band electronic excitations with energy transfer $\omega <2\mu$.
After integrating over all directions of the propagation of scattered photons, we find the spectral density of the angle-integrated Raman signal, 
\begin{equation}
g(\omega )\approx
\frac{1}{4}\Xi_{s}\left( \frac{e^{2}}{\pi \hbar c}\,\frac{v}{c}\right)^{2}
\frac{\omega}{\Omega^{2}} \theta(\omega-2\mu).
\label{eq:gmu}
\end{equation}
The total contribution of the electronic scattering to the total Raman efficiency is
\begin{equation}
I_{0}=\int_{0}^{\Omega/2}g(\omega )d\omega \sim 
\left( \frac{e^{2}}{2hc}\,\frac{v}{c}\right)^{2}\sim 10^{-10}.
\end{equation}
In doped graphene one may also expect to see some manifestation of the intra-band e-h excitations in the vicinity of the Fermi level and electron energies $\epsilon_{i}\approx\epsilon_{f}\approx\mu$.
Their analysis requires taking into account all terms in the expression for $R_{D}$, Eqs.~\eref{eq:RD}.
Then, we find that, for $\omega \leqslant (v/c)\Omega \ll \Omega <\gamma_{0}$, the yield of this low-energy feature is $\delta I=\int \delta g(\omega)d\omega \sim 10^{-15}$ for $\Omega \sim 1$eV.

\subsection{Raman scattering in bilayer graphene}
\label{sec:ramanbilayer}

For bilayer graphene with the interlayer coupling less then the base photon energy, $\gamma_{1}\ll\Omega,\tilde{\Omega}$, we also study the low energy excitations in the final states, in particular, those with $\omega\ll\gamma_{1}$.
The reason for restricting the following analysis to the low-energy excitations is that the higher-energy excitations decay fast due to electron-electron and electron-phonon processes, leaving very little chance to observe any structure in the Raman signal.
As a result, we employ the same expansion $\omega/\Omega\ll1$ as in section~\ref{sec:ramanmonolayer}.
Keeping terms up to the $\Omega^{-3}$ order [the latter  appear when the virtual state is taken to be in the high-energy bands] in the expansion of the Green's functions and performing summation over the intermediate states of the process, we get the amplitude $R$ in the form of a $4\times 4$ matrix
\begin{eqnarray}
R &\approx \frac{(e\hbar v)^{2}}{\Omega^{2}} \left( 
-i
\left(
\begin{array}{cc}
\sigma_{z} & 0 \\
0 & \sigma_{z} 
\end{array}
\right)
\left( \mathbf{l}\times\tilde{\mathbf{l}}^{*} \right)_{z} + \frac{\mathcal{M}\cdot\mathbf{d}}{\Omega} \right),
\nonumber
\\ 
\mathcal{M} &=
\left(
\begin{array}{cc}
\gamma_{1}\sigma_{x}(\mathbf{e}_{z}\times\boldsymbol{\sigma})\sigma_{x} & \mathbf{M} \\
\mathbf{M} & 0
\end{array}
\right) . \nonumber
\end{eqnarray}
Since the electronic modes with the excitation energies $\omega<\gamma_{1}/2$, result from the electron-hole pairs in the low-energy bands described by low-energy Hamiltonian $H_{\textrm{eff}}$ in Eq.~\eref{2b_Hamiltonian}, we take only the part of $R$ which acts in that two-dimensional Hilbert space, leaving the terms in the lowest relevant order in $vp/\gamma_{1}\ll 1$ and $\gamma_{1}/\Omega\ll 1$.
This projection results in effective amplitude of inelastic two-phonon process,
\begin{equation}
R_{\textrm{eff}}\approx
\frac{e^{2}\hbar^{2}v^{2}}{\Omega^{2}}\left\{
-i\sigma_{z} ( \mathbf{l}\times\tilde{\mathbf{l}}^{*})_{z} + \frac{\gamma_{1}}{\Omega} \left[ \sigma_{x}d_{y} + \sigma_{y}d_{x} \right]
\right\}.
\label{eff_amp}
\end{equation}
It is important to stress that for $\Omega\gg\gamma_{1}$ $R_{\textrm{eff}}$ cannot be correctly obtained within a effective two-bands theory, Eq.~\eref{2b_Hamiltonian}, without the described above intermediate steps, though for $\Omega<\gamma_{1}$ one could formally use $H_{\textrm{eff}}$ and Eqs.~(\ref{eq:Adef},\ref{int_Hamiltonian},\ref{eq:diagramelem},\ref{eq:RDdef}) to describe inelastic scattering of far-infrared light.
Within these approximations the scattering probability
\begin{equation}
w \approx
e^{4}\hbar v^{2}
\frac{\gamma_{1}}{\Omega^{4}}\left\{ \Xi_{s} + \frac{\gamma_{1}^{2}}{2\Omega^{2}} \Xi_{o} \right\},
\end{equation}
retains the characteristic of monolayer graphene --- the crossed polarisation of in/out photons in the dominant mode~\cite{Mucha3}.
The angle-integrated spectral density of Raman scattering
\begin{equation}
g(\omega) =  \frac{1}{4}\left(\frac{e^{2}}{\pi\hbar c}\frac{v}{c}\right)^{2}\frac{\gamma_{1}}{\Omega^{2}}
\left\{ \Xi_{s} + \frac{\gamma_{1}^{2}}{2\Omega^{2}} \Xi_{o} \right\}\theta(\omega-2\mu)
\label{spectral_density} 
\end{equation}
reflects the constant density of states characteristic for the parabolic spectrum of the bilayers.
This is different from monolayer graphene, where $g(\omega)\propto\omega$, reflecting the linear energy dependence of the density of states~\cite{Kashuba}.

\section{Inter-Landau-level transitions in the Raman spectrum of graphene}

Upon the quantisation of electron states into Landau levels Raman scattering inevitably causes inter-Landau-level transitions, which, having fixed energies, manifest themselves on the electronic contribution to the Raman plot as a pronounced structure which can be used to detect their contribution experimentally~\cite{erexp,erexp2}

\subsection{Monolayer graphene}

\begin{figure}[tbp]
\centering
\includegraphics[width=.95\columnwidth]{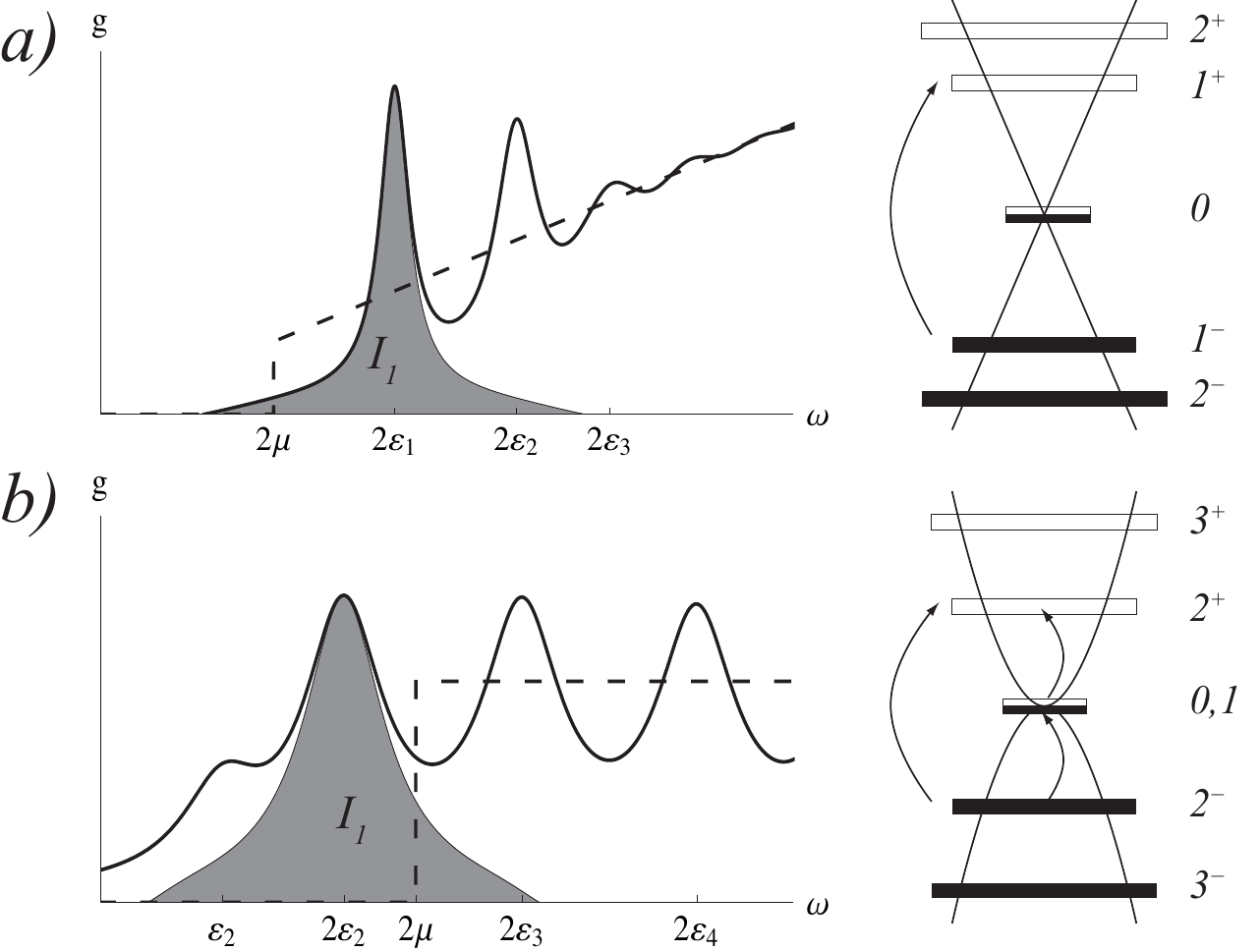}
\caption[Electron Raman spectra]{%
Spectral density $g(\omega )$ of light inelastically scattered from electronic excitations in undoped monolayer (a) and bilayer (b) graphene at quantising magnetic fields (solid line) and in doped graphene at $B=0$ (dashed line).
Sketch illustrates selection rules for processes corresponding to the visible features at Raman plot.
$\varepsilon_{n}$ denotes the LL energies in graphene.
In addition to dominant $n^{-}\to n^{+}$ processes at both plots, in bilayer spectrum the feature corresponding to $2^{-}\to 0$ and $0\to 2^{+}$ transitions can be recognised.}
\label{fig:graman}
\end{figure}

Electronic spectrum of monolayer graphene in a strong magnetic field can be described as a sequence $n^{\alpha}$ of Landau levels (LLs), corresponding to the states~\cite{McClure,AbergelFalko}
\begin{eqnarray}
|n^{\alpha }\rangle =\frac{1}{\sqrt{2}}
\left({ \Phi_{n} \atop i\alpha \Phi_{n-1}}\right)
\mbox{ for $n\geq 1$,}
\quad
|0\rangle =\left({\Phi _{0} \atop 0}\right),
\label{eq:llmono}
\\
\mbox{with energy $\alpha \varepsilon_{n}$, }
\varepsilon_{n}=\sqrt{2n}\,\hbar v/\lambda_{B},
\nonumber
\end{eqnarray}
where $\lambda_{B}=\sqrt{\hbar c/eB}$ is the magnetic length, $n$ enumerates the Landau levels, $\alpha=+$ denotes the conduction and $\alpha=-$ the valence band, and $\Phi_{n}$ are the normalised envelopes of LL wave functions.
Then, the corresponding elements of Feynman diagrams in Fig.~\ref{fig:diagrams} are
\begin{eqnarray*}
\raisebox{-12pt}{\begin{picture}(28,28)(-14,-14)
\thicklines
\put(-10,0){\line(1,0){20}}
\put(0,0){\vector(1,0){4}}
\end{picture}}&= G^{R/A}=\frac{\delta _{nn^{\prime }}\delta _{\alpha \alpha
^{\prime }}}{\varepsilon -\alpha \varepsilon _{n}\pm i0}, \\
\raisebox{-12pt}{\begin{picture}(28,28)(-14,-14)
\thicklines
\put(-8,12){\oval(8,8)[bl]}
\put(-8,4){\oval(8,8)[tr]}
\put(0,4){\oval(8,8)[bl]}
\put(-8,8){\vector(1,-1){4}}
\put(0,0){\line(1,1){12}}
\put(8,8){\vector(-1,-1){4}}
\put(0,0){\line(1,-1){12}}
\put(5,-5){\vector(1,-1){4}}
\put(0,0){\circle*{4}}
\end{picture}}&= \frac{ev\hbar }{2\sqrt{\Omega }}\mathbf{J}\cdot \mathbf{l}%
,\quad 
\raisebox{-12pt}{\begin{picture}(28,28)(-14,-14)
\thicklines
\put(-8,-12){\oval(8,8)[tl]}
\put(-8,-4){\oval(8,8)[br]}
\put(0,-4){\oval(8,8)[tl]}
\put(-4,-5){\vector(-1,-1){4}}
\put(0,0){\line(1,1){12}}
\put(8,8){\vector(-1,-1){4}}
\put(0,0){\line(1,-1){12}}
\put(5,-5){\vector(1,-1){4}}
\put(0,0){\circle*{4}}
\end{picture}}\,\,=\frac{ev\hbar }{2\sqrt{\Omega }}\mathbf{J}\cdot \mathbf{\tilde{l}}^{*}, \\
\mathbf{J}_{n^{\alpha }\,n^{\prime \alpha'}}& =
\alpha i\delta_{n',n-1}\mathbf{e}_{-}-\alpha'i\delta_{n'-1,n}%
\mathbf{e}_{+}, \\
R_{w} &= \frac{e^{2}v^{2}\hbar ^{2}}{6\gamma_{0}\Omega}
\xi\mathbf{J}\cdot \sum_{\pm }
\mathbf{e}_{\pm }(\mathbf{le}_{\mp })(\mathbf{\tilde{l}}^{*}\mathbf{e}_{\mp }).
\end{eqnarray*}%
Here $\mathbf{e}_{\pm }=\frac{1}{\sqrt{2}}\left( \mathbf{e}_{x}\pm i\mathbf{e}_{y}\right) $ is used to stress that a circularly polarised photon carries angular momentum $m=\pm 1$.
Matrix structure of the interaction terms allows the following selection rules for optically active inter-LL excitations in monolayer graphene:
\begin{equation}
\begin{array}{llll}
i)  & n^{-}\to n^{+}; & \multirow{2}{*}{iii)} & n^{-}\to (n+1)^{+}, 
\\
ii) & (n\mp 1)^{-}\to(n\pm 1)^{+}; & & (n+1)^{-}\to n^{+}.
\end{array}
\label{sel_rules}
\end{equation}
The excitation of the e-h pairs by Raman scattering in graphene at strong magnetic fields produces the electronic transition i) between LLs, with angular momentum transfer $\Delta m=0$ and excitation energy $\omega =2\varepsilon_{n}$, see Fig.~\ref{fig:graman}, and transitions ii), with $\Delta m=\pm 2$ and $\omega =\varepsilon_{n-1}+\varepsilon_{n+1}$.
The amplitudes of these two processes, 
\begin{eqnarray*}
R_{n^{-}\rightarrow n^{+}}=\frac{1}{4}\frac{(ev\hbar )^{2}}{c^{2}\Omega}
\sum_{\alpha =\pm }&\Biggl[
\frac{(\mathbf{le}_{+})(\mathbf{\tilde{l}}^{*}\mathbf{e}_{-})}{\Omega-\varepsilon_{n}-\alpha\varepsilon_{n+1}}
-
\frac{(\mathbf{le}_{+})(\mathbf{\tilde{l}}^{*}\mathbf{e}_{-})}{\varepsilon_{n}-\Omega-\alpha\varepsilon_{n-1}}
-\\&-
\frac{(\mathbf{le}_{-})(\mathbf{\tilde{l}}^{*}\mathbf{e}_{+})}{\Omega-\varepsilon_{n}-\alpha\varepsilon_{n+1}}
+
\frac{(\mathbf{le}_{-})(\mathbf{\tilde{l}}^{*}\mathbf{e}_{+})}{\varepsilon_{n}-\Omega-\alpha\varepsilon_{n-1}}\Biggr],
\end{eqnarray*}
\begin{eqnarray*}
R_{(n\mp 1)^{-}\rightarrow (n\pm 1)^{+}}=&\mp \frac{1}{4}\frac{(ev\hbar )^{2}}{c^{2}\Omega }
(\mathbf{le}_{\pm })(\mathbf{\tilde{l}}^{*}\mathbf{e}_{\pm })
\times \\ &\times
\sum_{\alpha =\pm }\Biggl[
\frac{\alpha}{\Omega-\varepsilon_{n+1}-\alpha\varepsilon_{n}}
+
\frac{\alpha }{\varepsilon_{n-1}-\Omega-\alpha\varepsilon_{n}}\
\Biggr],
\end{eqnarray*}
are such that $R_{n^{-}\to n^{+}}\gg R_{(n\mp 1)^{-}\to(n\pm 1)^{+}}$ for $\omega \ll \Omega $.
The latter relation is determined by a partial cancellation of the two diagrams constituting $R_{D}$.
Notice that the inter-LL modes $n^{-}\rightarrow n^{+}$ have the symmetry of the representation A$_{2}$ in Table~\ref{tab:list} and the same circular polarisation of ``in'' and ``out'' photons involved in its excitation.
Finally, the contact term $R_{w}$ in Fig.~\ref{fig:diagrams} allows for a weak transition iii) $n^{-}\to (n\pm 1)^{+}$, with the amplitude $R_{w}\ll R_{n^{-}\rightarrow n^{+}}$.
This transition, together with selection rules $\Delta m=\pm 1$ resembles the inter-LL transition involved in the far-infrared (FIR) absorption~\cite{Potemski2,AbergelFalko}.
However, the FIR-active excitation is 'valley-symmetric' \cite{AbergelFalko,Mucha2} and belongs to the representation E$_{1}$, whereas the Raman-active $n^{-}\to (n\pm 1)^{+}$ mode belongs to E$_{2}$, allowing the latter to couple to the $\Gamma$-point optical phonon and, thus, leading to the magneto-phonon resonance feature in the Raman spectrum \cite{Kechedzhi1}.
Also, $R_{w}$ originates from the trigonal warping term in $\mathcal{H}$ which violates the rotational symmetry of the Dirac Hamiltonian by transferring angular momentum $\pm 3$ from electrons to the lattice, allowing thus the change of angular momentum by $\pm1$.

\begin{table}[tbp]
\centering
\begin{tabular}{|c|c|c|c|}
\hline
C$_{6v}$ rep & transition & intensity & polarisation \\ \hline
E$_{2}$ & ${n^{-}\rightarrow (n+1)^{+} \atop (n+1)^{-}\rightarrow n^{+}}$ & %
\begin{minipage}{4.9cm} weak in Raman, strong in magneto-phonon resonance
\end{minipage} & $\sigma ^{\pm }\rightarrow \sigma ^{\mp }$ \\ \hline
A$_{1}$ & ${(n-1)^{-}\rightarrow (n+1)^{+} \atop (n+1)^{-}\rightarrow
(n-1)^{+}}$ & weak in Raman & $\sigma ^{\pm }\rightarrow \sigma ^{\mp }$ \\ 
\hline\hline
A$_{2}$ & {\scriptsize $n^{-}\rightarrow n^{+}$} & dominant in Raman & $%
\sigma ^{\pm }\rightarrow \sigma ^{\pm }$ \\ \hline\hline
\end{tabular}%
\caption{Raman-active inter-LL excitations in graphene.}
\label{tab:list}
\end{table}

The dominant inter-LL transitions $n^{-}\rightarrow n^{+}$ determine the spectral density of light scattered from electronic excitations in graphene at high magnetic fields: 
\begin{equation}
g_{n^{-}\rightarrow n^{+}}(\omega )\approx
\Xi_{s}\left( \frac{v^{2}}{c^{2}}\frac{e^{2}/\lambda _{B}}{\pi \Omega }\right) ^{2}
\sum_{n\geq 1}\gamma_{n}(\omega-2\varepsilon_{n}).
\label{eq:nntransition}
\end{equation}
Here we use Lorenzian $\gamma _{n}(x)=\pi ^{-1}\Gamma _{n}/[x^{2}+\Gamma _{n}^{2}]$, and inelastic LL broadening $\Gamma _{n}$ which increases with the LL number, and the factor $\Xi_{s}$ in Eq.~\eref{eq:nntransition} indicates that ``in'' and ``out'' photons have the same circular polarisation.

\subsection{Bilayer graphene}

Same as in non-magnetic case, e-h excitations in bilayer graphene are created in low-energy Landau levels, as at high energies the Landau level broadening due to, for example, electron-phonon interaction, will smear out the LL spectrum.
In strong magnetic fields, low-energy Landau levels in bilayer graphene can be described~\cite{AbergelMcCannFalko} by
\begin{eqnarray}
|n^{\alpha}\rangle = \frac{1}{\sqrt{2}}\left({\Phi_{n} \atop \alpha\Phi_{n-2}}\right)
\mbox{ for $n\geq 2$,}
\quad
|0/1\rangle = \left({\Phi_{0/1} \atop 0}\right),
\label{eq:llbi}
\\
\mbox{with energy }
\epsilon_{n^{\alpha}}=2\alpha\frac{\hbar^{2}v^{2}}{\gamma_{1}\lambda_{B}^{2}}\sqrt{n(n-1)}.
\nonumber
\end{eqnarray}
In a neutral bilayer, all LLs have additional four-fold degeneracy (two due to the electron spin and two due to the valley).
The peculiar $n=0$ and $n=1$ LLs are degenerate at $\epsilon=0$ producing to an 8-fold degenerate level.
Projecting the effective transition amplitude $R_{\textrm{eff}}$ onto the eigenstates $|n^{\alpha}\rangle$ we find the electronic Raman spectrum.
This procedure leads to the same selection rules as in monolayer graphene, see Eqs.~\eref{sel_rules}.
Among those, $i)$ are the dominant transitions.
For a undoped bilayer, the angle-integrated spectral density $g(\omega)$ of Raman scattering in the strong magnetic field is equal to:
\begin{eqnarray}
g(\omega) \approx &\left( \frac{v^{2}}{c^{2}}\frac{e^{2}/\lambda _{B}}{\pi \Omega }\right) ^{2}
\Biggl\{
\Xi_{s} \sum_{n\geq 2}\gamma(\omega - 2\epsilon_{n^{+}})
+
\Xi_{o}\left( \frac{\gamma_{1}}{\Omega} \right)^{2}
\times \nonumber\\&\times
\Biggl(
\sum_{n=1,2}\gamma(\omega-\epsilon_{(n+1)^{+}}) + \frac{1}{2}\sum_{n\geq 3}\gamma(\omega-\epsilon_{(n+1)^{+}} - \epsilon_{(n-1)^{+}}) 
\Biggr)
\Biggl\}.
\label{raman_mag_field}
\end{eqnarray}
The first term describes the dominant contribution due to the $n^{-}\to n^{+}$ transitions, while the last two terms describe the spectral density of the $(n\mp 1)^{-}\to {(n\pm 1)}^{+}$ transitions.

Low-energy electronic contribution to the Raman plot is shown with a solid line in Fig.~\ref{fig:graman}b.
The dominant peaks are due to the $n^{-}\to n^{+}$ transitions.
The quantum efficiency of a single $n^{-}\to n^{+}$ peak in Fig.~\ref{fig:graman}b is similar to value $I_{1}$ in monolayer graphene~\cite{Kashuba}.

A weaker feature in Fig.~\ref{fig:graman}b corresponds to both $2^{-}\rightarrow 0$ and $0\rightarrow 2^{+}$ transitions and is visible as it is positioned to the left of the first $2^{-}\rightarrow 2^{+}$ peak.
The quantum efficiencies of the $(n\pm 1)^{-}\to (n\mp 1)^{+}$ transitions are smaller by the factor $\left(\frac{\gamma_{1}}{\Omega}\right)^{2}$ in comparison to the $n^{-}\to n^{+}$ transitions.
This is different from the monolayer graphene case, where the corresponding ratio between quantum efficiencies of $(n\pm 1)^{-}\to (n\mp 1)^{+}$ and $n^{-}\to n^{+}$ transitions is $\left(\frac{\omega}{\Omega}\right)^{2}$, much smaller than for the bilayer.

\section{Magneto-phonon resonance in homogeneously and inhomogeneously doped and strained graphene}

Single- and multiple-phonon-emission lines in the Raman spectrum of graphene and the influence of the electron-phonon coupling on the phonon spectrum have been investigated in great detail~\cite{GeimNovoselov1,CastroNeto2,Basko1_1,Basko1_2,Basko2,Ando,Kechedzhi1,Pinczuk1,Pinczuk2,FerrariBasko}.
The coupling of electromagnetic field and phonon excitations is realised via the excitation of the virtual e-h pair, with the same symmetry properties as the $\Gamma$-point optical phonon which is created upon the recombination of the pair.
Analysing all possible combinations of polarisation vectors one may see from Table~\ref{tab:Reps} that the amplitude of Raman scattering with excitation of the phonon mode is proportional to 
\begin{equation}
R = R_{G} \sum_{\nu}\mathbf{u}_{\nu}\cdot\mathbf{d}
\end{equation}
that describes quantised sublattice displacement
\begin{equation}
\mathcal{U}(\mathbf{r},t)=(\mathcal{U}_{A}-\mathcal{U}_{B})/\sqrt{2} =
\sum_{\mathbf{k},\nu}\frac{\hbar}{\sqrt{2M\omega_{0}}} \left( \mathbf{u}_{\nu}
c_{\nu,\mathbf{k}} e^{i(\mathbf{k}\mathbf{r}-\omega_{0}t)/\hbar} + h.c. \right),
\end{equation}
where $M$ is the mass of single carbon atom, $\mathbf{u}$ is a polarisation of the phonon, and expression for $R_{G}$ can be taken from Basko~\cite{Basko2}.
Raman efficiency then is proportional to 
\begin{equation}
g \propto \sum_{\upsilon\nu} d_{\upsilon}d_{\nu}^{*} \, \mathrm{Im}\tilde{D}^{A}_{\upsilon\nu},
\end{equation}
where $\tilde D$ is the phonon propagator renormalized by electron-phonon interaction.
The latter strongly affects the energy of the phonon under the conditions of its resonance mixing with the inter-LL electronic transitions, the effect known as magneto-phonon resonance.

We use both phenomenological approach and the tight-binding model to study the interaction of electrons with $\Gamma$-point optical phonons in case of strong magnetic field.
The phenomenology is suitable to describe the low-energy excitations of electrons in the vicinity of Dirac points ($K$-points), while tight-binding model is needed to perform the integration over the whole Brillouin zone~\cite{Basko2} when estimating the intensity of the G-phonon-line in Raman.
To discuss the electron-phonon interaction it is instructive to trace the electron Hamiltonian
\begin{equation}
H=H_{0}-\mathbf{Q}\cdot\mathcal{U}
\label{eq:H+ph}
\end{equation}
back to the tight-binding model
\begin{eqnarray}
H_{0} &= \gamma_{0} \sum_{s}
\left(\begin{array}{cc}
0 & e^{i \mathbf{k}\mathbf{r}_{s}} \\
e^{-i \mathbf{k}\mathbf{r}_{s}} & 0
\end{array}\right)
\stackrel{\mathbf{k\to\xi K+p}}{\longrightarrow}
\xi v \, \boldsymbol{\sigma}\mathbf{p},
\\
\mathbf{Q} &= \sqrt{6}\frac{d \gamma_{0}}{d r}
\sum_{s}
\left(\begin{array}{cc}
0 & e^{i \mathbf{k}\mathbf{r}_{s}} \\
e^{-i \mathbf{k}\mathbf{r}_{s}} & 0
\end{array}\right)
\frac{\mathbf{r}_{s}}{a}
\stackrel{\mathbf{k\to\xi K+p}}{\longrightarrow}
F \, \boldsymbol{\sigma}\times\mathbf{e}_{z},
\end{eqnarray}
where $\mathbf{k}$ is a momentum within the Brillouin zone, $\mathbf{K}= (-\frac{4\pi}{3a},0)$, $v=\frac{\sqrt 3}{2} \gamma_{0} a /\hbar$, vectors $\mathbf{r}_1=(-\frac{a}{2},\frac{a}{2\sqrt3})$, $\mathbf{r}_2=(\frac{a}{2},\frac{a}{2\sqrt3})$, and $\mathbf{r}_3=(0,-\frac{a}{\sqrt3})$ are the positions of nearest ``B'' atoms with respect to ``A'' atom, and $F=\frac{3}{\sqrt 2}\,\frac{\partial \gamma_{0}}{\partial r}$, where $\frac{\partial \gamma_{0}}{\partial r}$ is a response of the hopping amplitude to the variation of the bond length.

The renormalisation of the phonon energy by its interaction with electrons is taken into account in the random phase approximation.
Defining electrons' bare Green's functions phonon bare propagator, and electron-phonon interaction vertex,
\begin{eqnarray*}
\raisebox{-11pt}{
\begin{picture}(28,28)(-14,-14)
\thicklines
\put(-8,0){\line(1,0){16}}
\put(0,0){\vector(1,0){4}}
\end{picture}}
&=\cases
{
G^{R/A} = (\varepsilon - H \pm i0)^{-1},
\atop\quad 
G^{K}   = (1-2f)(G^{R}-G^{A}),
}
\\
\raisebox{-11pt}{%
\begin{picture}(28,28)(-14,-14)
\thicklines
\put(-8,0){\line(1,0){4}}
\put(-2,0){\line(1,0){4}}
\put(4,0){\line(1,0){4}}
\end{picture}}
&= D^{A}_{\upsilon\nu} = \frac{2 \omega_0}{(\omega- i\gamma_{ph})^2-\omega_0^2}\delta_{\upsilon\nu},
\\
\raisebox{-11pt}{%
\begin{picture}(28,28)(-14,-14)
\thicklines
\put(-14,0){\line(1,0){4}}
\put(-9,0){\line(1,0){4}}
\put(3,3){\line(1,1){10}}
\put(9,9){\vector(-1,-1){4}}
\put(3,-3){\line(1,-1){10}}
\put(8,-8){\vector(1,-1){4}}
\put(0,0){\circle{8}}
\end{picture}}
&= -\frac{1}{2} \hbar a \sqrt{\frac{\sqrt 3}{M\omega_{0}}}\, \mathbf{Q},
\end{eqnarray*}
we use Dyson equation to find the renormalised phonon propagator:
\begin{eqnarray}
&\lambda\Pi_{\upsilon\nu}^A
\,=\,
\raisebox{-9pt}{\includegraphics[scale=.5]{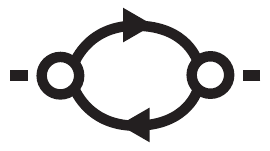}}
\,=\,
-i  \frac{\sqrt 3}{4}\frac{\hbar^{2}a^{2}}{M\omega_{0}} \int \frac{d\varepsilon}{2\pi}\int_{BZ} \frac{d^{2} \mathbf{p}}{(2\pi\hbar)^{2}}
\times\nonumber\\&\qquad\times
\mathrm{Tr}\Bigl\{
Q_{\upsilon} G^R(\varepsilon) Q_{\nu} G^K(\varepsilon+\omega)
+
Q_{\upsilon} G^K(\varepsilon) Q_{\nu} G^A(\varepsilon+\omega) \Bigr\},
\label{eq:PiDysona}
\\
&\quad\,\,\tilde{D}
\,=\,
\raisebox{-40pt}{\includegraphics[scale=.5]{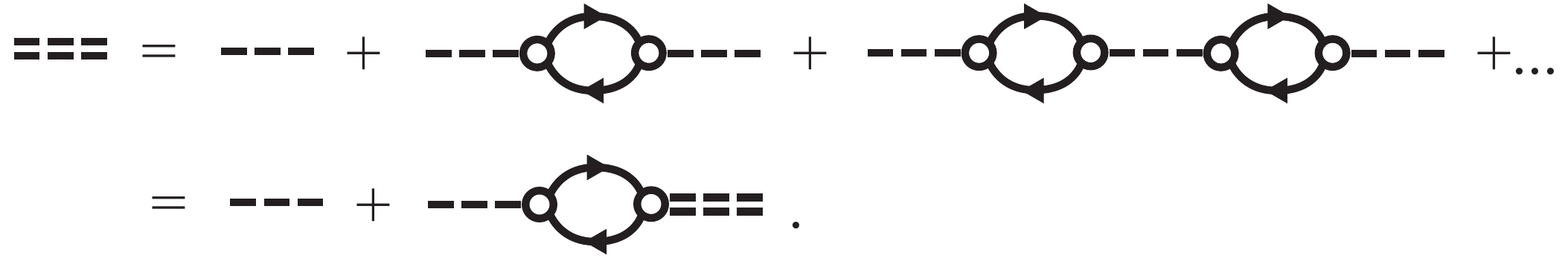}}
\label{eq:PiDysonb}
\end{eqnarray}
Here $\lambda =\frac{\sqrt 3}{M\omega_0}\left(\frac{Fa}{2v}\right)^2= \frac{3\sqrt 3}{2M\omega_0}\left(\frac{\hbar}{\gamma_0}\frac{d \gamma_0}{d r}\right)^{2}\approx 0.018$ is dimensionless constant of the electron-phonon coupling~\cite{Basko1_1}.
Trace is taken in $AB$ sites space and a summation over spins is already performed.
As momentum $k$ of a phonon created by the Raman process is small by parameter $v/c$, in the following it can be neglected.

The integral for polarisation loop in Eq.~\eref{eq:PiDysona} is divergent in the frame of linear Dirac model, so it can be split into two parts:
\begin{equation}
\Pi_{\upsilon\nu}^A = \Pi_{0} \delta_{\upsilon\nu} + \Delta\Pi_{\upsilon\nu},
\qquad 
\Pi_{0}\delta_{\upsilon\nu}=\Pi_{\upsilon\nu}^A|_{\omega,\mu,B\to0}\approx - 0.64\gamma_{0}\delta_{\upsilon\nu}.
\end{equation}
Here $\Pi_{0}$ is calculated numerically within tight-binding model, and $\Delta\Pi_{\upsilon\nu}$ is determined within the effective Dirac model, after subtracting the divergent part from the integrand.
For $B=0$ the answer is $\Delta\Pi_{\upsilon\nu}^{A} \propto \delta_{\upsilon\nu}$ and it repeats the derivation of Raman line shape found in Ref.~\cite{Ando,CastroNeto2,Ferrari2,Pinczuk1,DasSarma,Kechedzhi1,FerrariBasko}
\begin{equation*}
g = \Xi_{o} \, I_{G} \frac{\frac{1}{\pi}\gamma_{G}}{(\omega-\omega_{G})^{2}+\gamma_{G}^{2}},
\end{equation*}
where the width of the G-peak and its position are determined by
\begin{eqnarray}
\gamma_{G}=\gamma_{ph} - \frac{\lambda}{4} \tilde{\omega}_{0} \, \theta(\tilde{\omega}_{0}-2\mu),
\\
\omega_{G}=\tilde{\omega}_{0} - \lambda\frac{\mu}{\pi} + \lambda\frac{\tilde{\omega}_{0}}{4\pi} \left( \log\left|\frac{2\mu+ \tilde{\omega}_{0}}{2\mu-\tilde{\omega}_{0}}\right|  \right)
\Bigl),
\end{eqnarray}
where $\tilde{\omega}_{0}=\omega_0 + 0.64\lambda\gamma_{0}$ (we implied that $\omega_{0}\gg 0.64\lambda\gamma_{0} \gg \lambda \varepsilon_{F}$), and expression for $I_{G}$ can be taken from Basko~\cite{Basko2}.

\subsection{Magneto-phonon resonance in a homogeneously doped monolayer}

\begin{figure}
\centering
\includegraphics[width=.95\columnwidth]{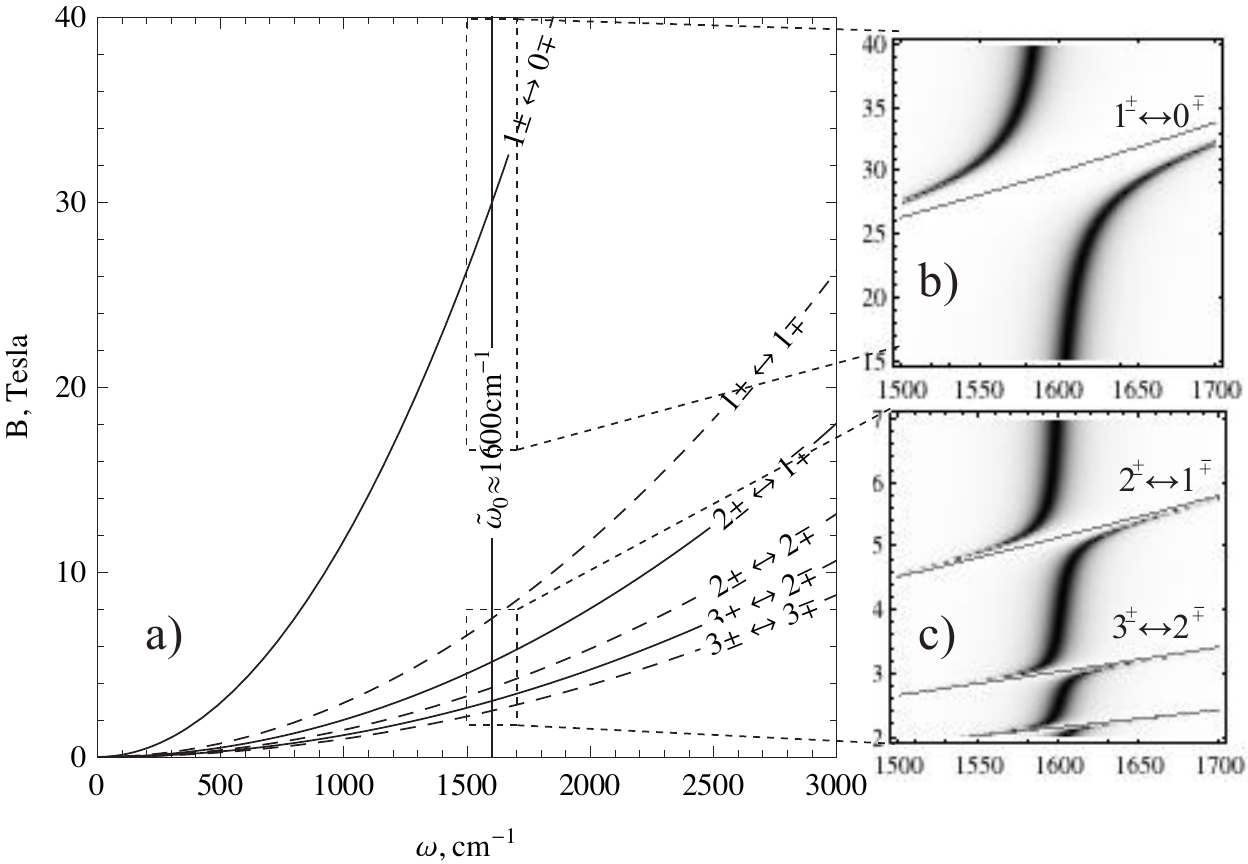}
\caption{%
G-peak splitting in magnetic field.
Density plots represent the dependence of the Raman efficiency on Raman shift and magnetic field.
a) Schematic plot shows phonon line, dashed curves represents dominant Raman active transitions $n^{-}\to n^{+}$, which do not interact with phonon line, and solid curves are weakly coupled Raman active transition $n^{-}\to(n\pm1)^{+}$, which interact with phonons forming the anticrossings;
b) first anticrossing due to the transitions $1^{-}\to 0$, $0\to 1^{+}$; 
c) 2$^{\rm nd}$, 3$^{\rm rd}$, and part of 4$^{\rm th}$ anticrossings.
In b) and c) plots are drawn for a zero electron density, so that the contributions from different polarisations are equal.}
\label{fig:magscheme}
\end{figure}

In the presence of magnetic field the continuous electronic spectra splits into the set of discrete Landau levels (LL).
Working in the basis of LLs Eqs.~(\ref{eq:llmono},\ref{eq:PiDysona},\ref{eq:PiDysonb}) and choosing the circularly polarized basis $\mathbf{e}_{\circlearrowright/\circlearrowleft}= \frac{1}{\sqrt 2}(\mathbf{e}_{x} \pm i \mathbf{e}_{y})$ for incoming and scattered light, we find that
\begin{equation*}
\Pi_{\upsilon\nu} =
(\mathbf{e}_{\circlearrowright})_{\upsilon}(\mathbf{e}_{\circlearrowright})_{\nu}^{*}
\Pi_{\circlearrowright}
+
(\mathbf{e}_{\circlearrowleft})_{\upsilon}(\mathbf{e}_{\circlearrowleft})_{\nu}^{*}
\Pi_{\circlearrowleft},
\mbox{ where }
\Pi_{\circlearrowright/\circlearrowleft}=\Pi_{0}+\Delta\Pi_{\circlearrowright/\circlearrowleft}.
\end{equation*}
For the chemical potential pinned at the zero LL
\begin{equation}
\Delta\Pi_{\circlearrowright/\circlearrowleft} =
\frac{\varepsilon_{1}^{2}}{\pi} \Biggl[
\frac{-f_{0}}{\varepsilon_{1}\pm \omega}
+
\frac{1-f_{0}}{-\varepsilon_{1}\pm \omega}
+
\frac{1}{\varepsilon_{1}}
-\omega^{2}
\sum_{n\geq 2}^{\infty}
\frac{1}{\Omega_{n,+}(\Omega_{n,+}^{2}-\omega^{2})}
\Biggr],
\end{equation}
where $f_{0}$ is a density-dependent filling factor of the zero level and $\Omega_{n,\pm}=\varepsilon_{n}\pm\varepsilon_{n-1}$ are the frequencies of electron-hole excitations.
If $\mu$ is large enough, so that $m^{\textrm{th}}$ level ($m\geq 1$) is partially filled, then
\begin{eqnarray}
\Delta\Pi_{\circlearrowright/\circlearrowleft} =
\frac{\varepsilon_{1}^{2}}{2\pi} \Biggl[
\frac{f_{m^+}-1}{\Omega_{m,-}\pm \omega}
+
\frac{-f_{m^+}}{\Omega_{m+1,-}\pm \omega}
+
\frac{f_{m^+}-1}{\Omega_{m,+}\pm \omega}
+\nonumber\\+
\frac{1-f_{m^+}}{-\Omega_{m+1,+}\pm \omega}
+
\frac{-1}{\Omega_{m,+}\pm \omega}
+
2\frac{\varepsilon_{m+1}}{\varepsilon_{1}^{2}}
-2\omega^{2}
\sum_{n\geq m+2}^{\infty}
\frac{1}{\Omega_{n,+}(\Omega_{n,+}^{2}-\omega^{2})}
\Biggr].
\label{eq:deltaPiB}
\end{eqnarray}
Then, the net Raman efficiency of the $\Gamma$-point phonon mode in graphene in a strong magnetic field is
\begin{equation}
g =
\sum_{\mathcal{A}= \circlearrowleft/\circlearrowright}|\mathbf{d}\mathbf{e}_{\mathcal{A}}|^{2}  
\frac{I_{G}\frac{1}{\pi}\gamma_{ph}}{(\omega-\tilde{\omega}_{0}-\lambda\,\Delta\Pi_{\mathcal{A}}(\omega))^{2}+\gamma_{ph}^{2}},
\label{eq:gdeltaPiB}
\end{equation}
where the two terms in the sum correspond to the opposite circular polarisations of the optical phonon, and it is plotted in Fig.~\ref{fig:magscheme} for several doping densities.

The line described by the Eq.~\eref{eq:gdeltaPiB} displays two types of typical behaviour depending on the range of parameters.
Far from the resonance only a slight shift of the electron and phonon excitations exists,
\begin{equation}
\omega_{\circlearrowleft/\circlearrowright} \approx
\tilde{\omega}_{0} + \lambda \, \Delta\Pi_{\circlearrowleft/\circlearrowright}(\tilde{\omega}_{0}).
\end{equation}
Since the line shift may differ for the opposite circular polarisations of the phonon, this leads to a small splitting of the G-line in Raman.
In the vicinity of the intersection of the phonon line with the electron-hole excitation line for several doping densities the polarisation operator becomes large and causes significant change in spectra, forming an anticrossing, in addition to the line splitting.
The form of the line is determined by four solutions $\omega_{\circlearrowleft/\circlearrowright}^{\pm}$
\begin{equation}
\omega - \tilde{\omega}_{0} - \lambda\,\Delta\Pi_{\circlearrowleft/\circlearrowright}(\omega)=0,
\end{equation}
two for each polarisation with energies above ($+$) or below ($-$) the bare phonon line.
The form of such anticrossings in undoped graphene is shown at Fig.~\ref{fig:magscheme}.
The resonance takes more sophisticated form in the case of the doped graphene.
With increasing of the electron density (and, correspondingly, the filling factor), the transition $1^{-}\to0^{+}$ invoked by anticlockwise polarised light is weaken by Pauli blocking (see first anticrossing at Fig.~\ref{fig:mrplotsdensity}a, while the $0^{-}\to1^{+}$ transition remains.
When the $n=0$ Landau level is fully occupied ($n_{e}=1/\pi \lambda_{B}^{2} \approx 1.4\times 10^{12}\,\mbox{cm}^{-12}$,), in anticlockwise polarisation the anticrossing completely blocked (see Fig.~\ref{fig:mrplotsdensity}b, which coincides with the experimentally observed features~\cite{erexp}.

\subsection{Influence of inhomogeneous doping and strain on the magneto-phonon resonance lines shape}

In graphene deposited on the substrate, the density of the electrons is very often, not the same everywhere in the sample, but varies, forming puddles at the length of $100\div500nm$~\cite{Klitzing1}.
Also, ripples of graphene sheet, induce inhomogeneous strain.
At weak magnetic fields the the electron density and strain disorder lead to the scattering of the charge carriers.
The situation changes when the external magnetic field ($\sim25T$) is applied.
At such a strong field, the magnetic length is about $5nm$, which is much smaller than the typical size of the spatial fluctuations of both carriers density and strain.
In this case, the Raman response of the graphene flake is the average of the locally defined spectra.
Below, we discuss how such averaging changes the fine structure of the magneto-phonon resonance.

\paragraph{Inhomogeneous density}

\begin{figure}
\centering
\includegraphics[width=.95\columnwidth]{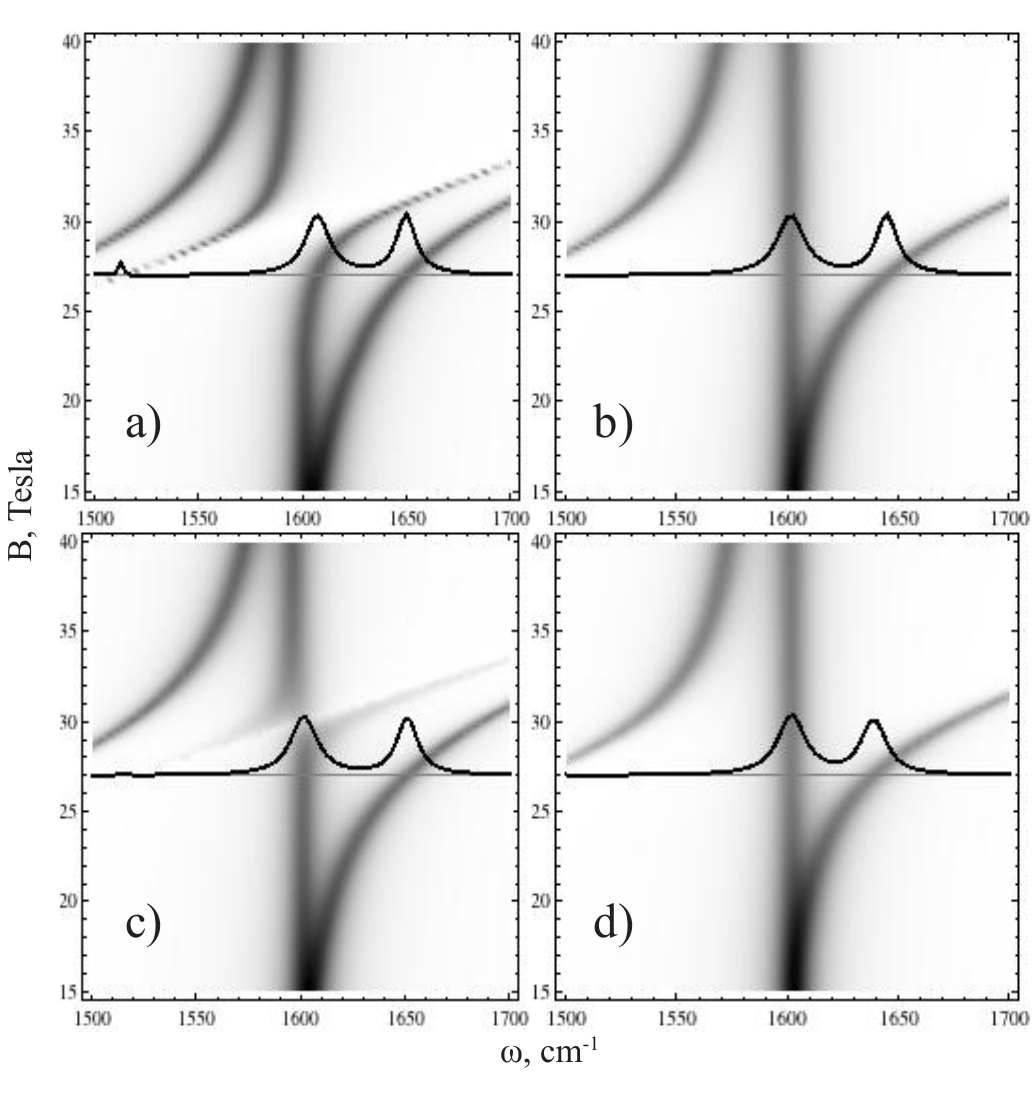}
\caption{%
Anticrossing of $1^{-}\to0^{+}$ and $0^{-}\to1^{+}$ transition with the optical phonon for various levels of graphene doping.
a) homogeneous graphene, electron density $n_{e}=1\times10^{12}\,\mbox{cm}^{-2}$: transitions in both circular polarisations are active, but $1^{-}\to 0^{+}$ is suppressed;
b) graphene with a homogeneous doping with electron density $n_{e}=2\times10^{12}\,\mbox{cm}^{-2}$: only $0^{-}\to 1^{+}$ transition is active;
c) graphene with inhomogeneous electron density, such that the average density is $\bar{n}_{e}=1.5\times10^{12}\,\mbox{cm}^{-2}$ and distribution with the variance $\delta n_{e} = 0.4\times10^{12}\,\mbox{cm}^{-2}$.
d) The same as in c), but with $\bar{n}_{e}=2.3\times10^{12}\,\mbox{cm}^{-2}$ and $\delta n_{e} = 0.4\times10^{12}\,\mbox{cm}^{-2}$.}
\label{fig:mrplotsdensity}
\end{figure}

The spatial distribution of the electrons density, results in summing up of the local spectra contributions from the various parts of the graphene flake.
The net picture is the overlay of the anticrossings at different filling factors (like one shown at Fig.~\ref{fig:mrplotsdensity}a,b) weighted with the normal distribution factor, which we use mostly for the illustration purposes.

As shown in Fig.~\ref{fig:mrplotsdensity}c,d, this leads to the inhomogeneous broadening of the side lines related to the variation of the size of the anticrossing gaps between the $0^{-}\to1^{+}$ transition and the phonon, and it blurs out the much weaker splitting/anticrossing of the phonon and $1^{-}\to0^{+}$ transition.

\paragraph{Inhomogeneously strained graphene}

\begin{figure}
\centering
\includegraphics[width=.95\columnwidth]{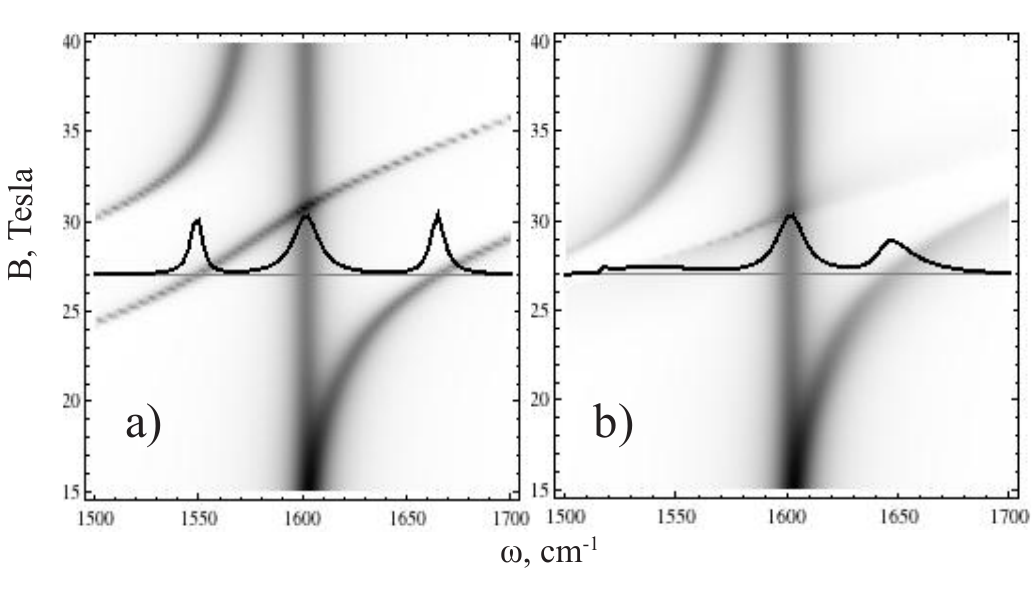}
\caption{%
The $0^{-}\to1^{+}$ and $1^{-}\to0^{+}$ anticrossing for strained graphene with electron density $n_{e}=2\times10^{12}\,\mbox{cm}^{-2}$.
a) homogeneously strained graphene with effective $B_\mathrm{strain}=3$\,T: one anticrossing splits into two shifted vertically by $2B_\mathrm{strain}$;
b) graphene subjected to inhomogeneous strain averaged over the distribution with the variance $\delta B_\mathrm{strain}=3$\,T: the peak corresponding to $0^{-}\to1^{+}$ transition has a distinctive asymmetric shape.}
\label{fig:mrplotsstrain}
\end{figure}


Inhomogeneous strain in graphene is often associated with magnetic field, $B_\mathrm{strain}$ for electrons in $K$ valley and $-B_\mathrm{strain}$ in $K'$, which estimated as $B_\mathrm{strain}\sim \frac{\hbar c}{e a} \frac{h^{2}}{l^{3}}$, where $a$ is the bond length, $l$ and $h$ are the linear scale and height of a ripple in graphene~\cite{RMPgraphene}.
In the external magnetic field $B$ the total effective field acting on electrons in the valleys $K$ and $K'$ is different,
\begin{equation*}
B_{\xi}=B+\xi B_\mathrm{strain},
\end{equation*}
where $\xi=+$ for $K$ point and $\xi=-$ for $K'$.
Difference in magnetic field leads to the different Landau spectra in each valley, thus, splitting the values of the external field B, for which, e.g.~the resonance of $0^{-}_{K}\to1^{+}_{K}$ and $0^{-}_{K'}\to1^{+}_{K'}$ with the phonon is realised.
Formally, this can be taken into account by a substitution $\Delta \Pi(B)\to\frac{1}{2}\sum_{\xi=\pm}\Delta\Pi(B_{\xi})$ in Eq.~\eref{eq:gdeltaPiB}.

For the local Raman spectrum related to an area with a fixed $B_{strain}$, this results in a fine structure of the anticrossing shown in Fig.~\ref{fig:mrplotsstrain}a, where an additional line appears seemingly crossing the phonon mode without splitting.
Such a behaviour occurs due to the fact that instead of an anticrossing/mixing of two modes, now, three modes are coupled, forming one phonon-electron mixed excitation with energy pinned in the middle between the two strongly split side bands.
Here, the central phonon line corresponds to the circular polarisation of the phonon for which mixing with electronic modes is suppressed by Pauli blocking, due to high density of electrons used in the calculation.
Averaged out a normal distribution of $B_\mathrm{strain}$ with zero average expectation and variance $\delta B_\mathrm{strain}=3$T, the anticrossing branches become asymmetrically broadened, as shown in Fig.~\ref{fig:mrplotsstrain}b.

\section{Summary}

We presented a theory of inelastic scattering of photons in mono- and bilayer graphene accompanied by the excitation of electron-hole pairs.
The dominant scattering processes are characterised by the crossed polarisation of in/out photons while the Raman efficiency is proportional to the density of states (linear increase with Raman shift for monolayer and constant for bilayer) and in case of doping the gap of width $2\mu$ opens up.

In strong magnetic field Raman spectra acquires the pronounced structure with peaks corresponding to the inter-Landau-level transitions.
The selection rules for the strongest mode preserve LL number $n$: $n^{-}\to n^{+}$, and quantum efficiency of the single peak in Fig.~\ref{fig:graman} is $I_{1}\sim \left( \frac{v^{2}}{c^{2}}\,\frac{e^{2}/\lambda _{B}}{\pi \Omega }\right)^{2}\sim 10^{-12}$ per incoming photon for $\Omega\sim 1$eV photons in magnetic field $B\sim 10$T.
The theoretical prediction for the intensity of the phonon-induced $G$ peak,\cite{Basko2} which is a well known Raman feature in carbon materials\cite{Ferrari2}, estimates $I_{G}\sim 10^{-11}$.
This result is close in the order of magnitude to the intensity of a single $n^{-}\to n^{+}$ peak.
So, the spectral features of inter-Landau-level transitions in graphene can be observed experimentally~\cite{erexp}.
The weaker mode with same selection rules, similar to the absorption process, $n^{-}\to(n\pm1)^{+}$, but belonging to different symmetry group representation (E$_{2}$, while absorption has E$_{1}$), is coupled to the phonon mode, creating the anticrossing and the G-peak line.

\ack
We thank A.~Ferrari, D.~Smirnov, M.~Mucha-Kruczynski, K.~Kechedzhi, M.~Goerbig, F.~Guinea, T.~Ando, and D.~Basko for useful discussions. This reviewed work was supported by EPSRC, the Royal Society, EC ERC, Royal Society Wolfson Research Merit Award, and European Research Council Advance Investigator grant `Graphene and Beyond'.

\section*{References}

\bibliographystyle{iopart-num}
\bibliography{G-Raman}

\providecommand{\newblock}{}
\begin{thebibliography}{10}
\expandafter\ifx\csname url\endcsname\relax
  \def\url#1{{\tt #1}}\fi
\expandafter\ifx\csname urlprefix\endcsname\relax\def\urlprefix{URL }\fi
\providecommand{\eprint}[2][]{\url{#2}}

\bibitem{Ohta}
Ohta T, Bostwick A, Seyller T, Horn K and Rotenberg E 2006 {\em Science\/} {\bf
  313} 951

\bibitem{Mucha1}
Mucha-Kruczy\'{n}ski M, Tsyplyatyev O, Grishin A, McCann E, Fal'ko V~I,
  Bostwick A and Rotenberg E 2008 {\em Phys. Rev. B\/} {\bf 77}(19) 195403

\bibitem{Potemski2}
Sadowski M~L, Martinez G, Potemski M, Berger C and de~Heer W~A 2006 {\em Phys.
  Rev. Lett.\/} {\bf 97}(26) 266405

\bibitem{Kim2}
Jiang Z, Henriksen E~A, Tung L~C, Wang Y~J, Schwartz M~E, Han M~Y, Kim P and
  Stormer H~L 2007 {\em Phys. Rev. Lett.\/} {\bf 98}(19) 197403

\bibitem{Kuzmenko1}
Kuzmenko A~B, van Heumen E, van~der Marel D, Lerch P, Blake P, Novoselov K~S
  and Geim A~K 2009 {\em Phys. Rev. B\/} {\bf 79}(11) 115441

\bibitem{Basov}
Zhang L~M, Li Z~Q, Basov D~N, Fogler M~M, Hao Z and Martin M~C 2008 {\em Phys.
  Rev. B\/} {\bf 78}(23) 235408

\bibitem{AbergelFalko}
Abergel D~S~L and Fal'ko V~I 2007 {\em Phys. Rev. B\/} {\bf 75}(15) 155430

\bibitem{Geim}
Blake P, Hill E~W, Neto A~H~C, Novoselov K~S, Jiang D, Yang R, Booth T~J and
  Geim A~K 2007 {\em Applied Physics Letters\/} {\bf 91} 063124

\bibitem{AbergelRussellFalko}
Abergel D~S~L, Russell A and Fal'ko V~I 2007 {\em Applied Physics Letters\/}
  {\bf 91} 063125

\bibitem{Kuzmenko2}
Kuzmenko A~B, Crassee I, van~der Marel D, Blake P and Novoselov K~S 2009 {\em
  Phys. Rev. B\/} {\bf 80}(16) 165406

\bibitem{Kim1}
Henriksen E~A, Jiang Z, Tung L~C, Schwartz M~E, Takita M, Wang Y~J, Kim P and
  Stormer H~L 2008 {\em Phys. Rev. Lett.\/} {\bf 100}(8) 087403

\bibitem{Kim3}
Li Z~Q, Henriksen E~A, Jiang Z, Hao Z, Martin M~C, Kim P, Stormer H~L and Basov
  D~N 2009 {\em Phys. Rev. Lett.\/} {\bf 102}(3) 037403

\bibitem{Heinz}
Mak K~F, Lui C~H, Shan J and Heinz T~F 2009 {\em Phys. Rev. Lett.\/} {\bf
  102}(25) 256405

\bibitem{Kuzmenko3}
Kuzmenko A~B, Benfatto L, Cappelluti E, Crassee I, van~der Marel D, Blake P,
  Novoselov K~S and Geim A~K 2009 {\em Phys. Rev. Lett.\/} {\bf 103}(11) 116804

\bibitem{GeimNovoselov3}
Nair R~R, Blake P, Grigorenko A~N, Novoselov K~S, Booth T~J, Stauber T, Peres
  N~M~R and Geim A~K 2008 {\em Science\/} {\bf 320} 1308

\bibitem{Gaskell}
Gaskell P~E, Skulason H~S, Rodenchuk C and Szkopek T 2009 {\em Applied Physics
  Letters\/} {\bf 94} 143101

\bibitem{Ferrari}
Ferrari A~C, Meyer J~C, Scardaci V, Casiraghi C, Lazzeri M, Mauri F, Piscanec
  S, Jiang D, Novoselov K~S, Roth S and Geim A~K 2006 {\em Phys. Rev. Lett.\/}
  {\bf 97} 187401

\bibitem{Graf}
Graf D, Molitor F, Ensslin K, Stampfer C, Jungen A, Hierold C and Wirtz L 2007
  {\em Nano Letters\/} {\bf 7} 238--242

\bibitem{CastroNeto1}
Malard L~M, Nilsson J, Elias D~C, Brant J~C, Plentz F, Alves E~S, Castro~Neto
  A~H and Pimenta M~A 2007 {\em Phys. Rev. B\/} {\bf 76}(20) 201401

\bibitem{Jiang}
Jiang J~W, Tang H, Wang B~S and Su Z~B 2008 {\em Phys. Rev. B\/} {\bf 77}(23)
  235421

\bibitem{Potemski1}
Faugeras C, Nerri\`{e}re A, Potemski M, Mahmood A, Dujardin E, Berger C and
  de~Heer W~A 2008 {\em Applied Physics Letters\/} {\bf 92} 011914

\bibitem{Berciaud}
Berciaud S, Ryu S, Brus L~E and Heinz T~F 2009 {\em Nano Letters\/} {\bf 9} 346

\bibitem{BalandinCalizo}
Calizo I, Bejenari I, Rahman M, Liu G and Balandin A~A 2009 {\em Journal of
  Applied Physics\/} {\bf 106} 043509

\bibitem{GeimNovoselov1}
Pisana S, Lazzeri M, Casiraghi C, Novoselov K~S, Geim A~K, Ferrari A~C and
  Mauri F 2007 {\em Nat Mater\/} {\bf 6} 198--201

\bibitem{CastroNeto2}
Castro~Neto A~H and Guinea F 2007 {\em Phys. Rev. B\/} {\bf 75} 045404

\bibitem{Basko1_1}
Basko D~M 2008 {\em Phys. Rev. B\/} {\bf 78} 125418

\bibitem{Basko1_2}
Basko D~M 2007 {\em Phys. Rev. B\/} {\bf 76} 081405

\bibitem{Ando}
Ando T 2007 {\em Journal of the Physical Society of Japan\/} {\bf 76} 024712

\bibitem{Kechedzhi1}
Goerbig M~O, Fuchs J~N, Kechedzhi K and Fal'ko V~I 2007 {\em Phys. Rev.
  Lett.\/} {\bf 99} 087402

\bibitem{Pinczuk1}
Yan J, Zhang Y, Kim P and Pinczuk A 2007 {\em Phys. Rev. Lett.\/} {\bf 98}(16)
  166802

\bibitem{Pinczuk2}
Yan J, Henriksen E~A, Kim P and Pinczuk A 2008 {\em Phys. Rev. Lett.\/} {\bf
  101}(13) 136804

\bibitem{FerrariBasko}
Basko D~M, Piscanec S and Ferrari A~C 2009 {\em Phys. Rev. B\/} {\bf 80} 165413

\bibitem{Wallace}
Wallace P~R 1947 {\em Phys. Rev.\/} {\bf 71}(9) 622--634

\bibitem{RMPgraphene}
Castro~Neto A~H, Guinea F, Peres N~M~R, Novoselov K~S and Geim A~K 2009 {\em
  Rev. Mod. Phys.\/} {\bf 81}(1) 109--162

\bibitem{GeimNovoselov2}
Geim A~K and Novoselov K~S 2007 {\em Nat Mater\/} {\bf 6} 183--191

\bibitem{McCann2}
McCann E and Fal'ko V~I 2006 {\em Phys. Rev. Lett.\/} {\bf 96}(8) 086805

\bibitem{PlatzmanWolff}
Platzman P~M, Wolff P~A and Tzoar N 1968 {\em Phys. Rev.\/} {\bf 174} 489

\bibitem{PlatzmanWolff2}
Platzmann P~M and Wolff P~A 1973 {\em Waves and interactions in solid state
  plasmas\/} (New York: Academic Press)

\bibitem{Chinese1}
Hong-Yan L and Qiang-Hua W 2008 {\em Chinese Physics Letters\/} {\bf 25} 3746

\bibitem{McClure}
McClure J~W 1957 {\em Phys. Rev.\/} {\bf 108}(3) 612--618

\bibitem{Dresselhaus}
Saito R, Dresselhaus G and Dresselhaus M~S 1998 {\em Physical properties of
  carbon nanotubes\/} (London: Imperial College Press)

\bibitem{McCann1}
McCann E, Kechedzhi K, Fal'ko V~I, Suzuura H, Ando T and Altshuler B~L 2006
  {\em Phys. Rev. Lett.\/} {\bf 97}(14) 146805

\bibitem{Malard2}
Mafra D~L, Malard L~M, Doorn S~K, Htoon H, Nilsson J, Castro~Neto A~H and
  Pimenta M~A 2009 {\em Phys. Rev. B\/} {\bf 80}(24) 241414

\bibitem{Kechedzhi2}
Kechedzhi K, McCann E, Fal'ko V~I, Suzuura H, Ando T and Altshuler B~L 2007
  {\em The European Physical Journal - Special Topics\/} {\bf 148}(1) 39--54
  ISSN 1951-6355

\bibitem{Dresselhaus2}
Dresselhaus M and Dresselhaus G 1981 {\em Advances in Physics\/} {\bf 30}
  139--326

\bibitem{Chakraborty}
Das A, Chakraborty B, Piscanec S, Pisana S, Sood A~K and Ferrari A~C 2009 {\em
  Phys. Rev. B\/} {\bf 79}(15) 155417

\bibitem{Kashuba}
Kashuba O and Fal'ko V~I 2009 {\em Phys. Rev. B\/} {\bf 80}(24) 241404

\bibitem{Mucha3}
Mucha-Kruczy\'{n}ski M, Kashuba O and Fal'ko V~I 2010 {\em Phys. Rev. B\/} {\bf
  82}(4) 045405

\bibitem{erexp}
Faugeras C, Amado M, Kossacki P, Orlita M, K{\"u}hne M, Nicolet A~A~L, Latyshev
  Y~I and Potemski M 2011 {\em Phys. Rev. Lett.\/} {\bf 107} 036807

\bibitem{erexp2}
Kossacki P, Faugeras C, K\"{u}hne M, Orlita M, Nicolet A~A~L, Schneider J~M,
  Basko D~M, Latyshev Y~I and Potemski M 2011 {\em Phys. Rev. B\/} {\bf 84}(23)
  235138

\bibitem{Mucha2}
Mucha-Kruczy\'{n}ski M, Abergel D~S~L, McCann E and Fal'ko V~I 2009 {\em
  Journal of Physics: Condensed Matter\/} {\bf 21} 344206

\bibitem{AbergelMcCannFalko}
Abergel D~S, McCann E and Fal'ko V~I 2007 {\em European Physical Journal -
  Special Topics\/} {\bf 148}(1) 105

\bibitem{Basko2}
Basko D~M 2009 {\em New Journal of Physics\/} {\bf 11} 095011

\bibitem{Ferrari2}
Ferrari A~C 2007 {\em Solid State Communications\/} {\bf 143} 47--57

\bibitem{DasSarma}
Hwang E~H and Das~Sarma S 2007 {\em Phys. Rev. B\/} {\bf 75}(20) 205418

\bibitem{Klitzing1}
Martin J, Akerman N, Ulbricht G, Lohmann T, Smet J~H, von Klitzing K and Yacoby
  A 2008 {\em Nat. Phys.\/} {\bf 4} 144--148

\end{thebibliography}

\end{document}